\documentstyle[12pt,epsfig]{article}
 
\newskip\humongous \humongous=0pt plus 1000pt minus 1000pt

\newif\ifdtup

\def\abs#1{\left| #1\right|}
\def\pr#1{#1^\prime}

\def\beq{\begin{equation}}
\def\eeq{\end{equation}}

\def\beqn{\begin{eqnarray}}
\def\eeqn{\end{eqnarray}}
\relax

\def\dotx{\dotx{\dot\overline{x}}}

\relax

\catcode`\@=11
 
\@addtoreset{equation}{section}
\def\theequation{\thesection\arabic{equation}}
 
\def\@normalsize{\@setsize\normalsize{15pt}\xiipt\@xiipt
\abovedisplayskip 14pt plus3pt minus3pt%
\belowdisplayskip \abovedisplayskip
\abovedisplayshortskip \z@ plus3pt%
\belowdisplayshortskip 7pt plus3.5pt minus0pt}
 
\def\small{\@setsize\small{13.6pt}\xipt\@xipt
\abovedisplayskip 13pt plus3pt minus3pt%
\belowdisplayskip \abovedisplayskip
\abovedisplayshortskip \z@ plus3pt%
\belowdisplayshortskip 7pt plus3.5pt minus0pt
\def\@listi{\parsep 4.5pt plus 2pt minus 1pt
     \itemsep \parsep
     \topsep 9pt plus 3pt minus 3pt}}
 
\@twosidetrue
\relax

\catcode`@=12
 
\catcode`\@=11
 
\def\section{\@startsection{section}{1}{\z@}{3.5ex plus 1ex minus
   .2ex}{2.3ex plus .2ex}{\large\bf}}
 
\def\thesection{\arabic{section}.}

\def\appendix{\setcounter{section}{0}
 \def\thesection{APPENDIX \Alph{section}:}
 \def\theequation{\Alph{section}.\arabic{equation}}}
 
\def\ps@headings{\def\@oddfoot{}\def\@evenfoot{}
\def\@oddhead{\hbox{}\hfill
 \makebox[.5\textwidth]{\raggedright\ignorespaces --\thepage{}--
 \hfill {}}}  
\def\@evenhead{\@oddhead}
\def\subsectionmark##1{\markboth{##1}{}}
}
 
\ps@headings
 
\catcode`\@=12

\def\figcap{\section*{Figure Captions\markboth
 {FIGURECAPTIONS}{FIGURECAPTIONS}}\list
 {Fig. \arabic{enumi}:\hfill}{\settowidth\labelwidth{Fig. 999:}
 \leftmargin\labelwidth
 \advance\leftmargin\labelsep\usecounter{enumi}}}
 \relax
\def\tablecap{\section*{Table Captions\markboth
 {TABLECAPTIONS}{TABLECAPTIONS}}\list
 {Table \arabic{enumi}:\hfill}{\settowidth\labelwidth{Table 999:}
 \leftmargin\labelwidth
 \advance\leftmargin\labelsep\usecounter{enumi}}}
 \relax
\def\reflist{\section*{References\markboth
 {REFLIST}{REFLIST}}\list
 {[\arabic{enumi}]\hfill}{\settowidth\labelwidth{[999]}
 \leftmargin\labelwidth
 \advance\leftmargin\labelsep\usecounter{enumi}}}
 \relax
 
\catcode`\@=11
 
\def\ps@headings{\def\@oddfoot{}\def\@evenfoot{}
\def\@oddhead{\hbox{}\hfill
 \makebox[.5\textwidth]{\raggedright\ignorespaces --\thepage{}--
 \hfill {}}}    
\def\@evenhead{\@oddhead}
\def\subsectionmark##1{\markboth{##1}{}}
}
 
\ps@headings
 
\relax

\relax
\def\pl#1#2#3{{\it Phys. Lett. }{\bf #1}(19#2)#3}
\def\zp#1#2#3{{\it Z. Phys. }{\bf #1}(19#2)#3}
\def\prl#1#2#3{{\it Phys. Rev. Lett. }{\bf #1}(19#2)#3}

\def\pr#1#2#3{{\it Phys. Rev. }{\bf #1}(19#2)#3}
\def\np#1#2#3{{\it Nucl. Phys. }{\bf #1}(19#2)#3}

\relax

  \newcommand{\ccaption}[2]{
    \begin{center}
    \parbox{0.85\textwidth}{
      \caption[#1]{\small{\it{#2}}}
      }
    \end{center}
    }
\begin{document}
\def\theequation{\arabic{equation}}
\newcommand\sss{\scriptscriptstyle}
\newcommand\mug{\mu_\gamma}
\newcommand\mue{\mu_e}
\newcommand\muf{\mu_{\sss F}}
\newcommand\mur{\mu_{\sss R}}
\newcommand\muo{\mu_0}
\newcommand\me{m_e}
\newcommand\as{\alpha_{\sss S}}         
\newcommand\ep{\epsilon}
\newcommand\epb{\overline{\epsilon}}
\newcommand\aem{\alpha_{\rm em}}
\newcommand\refq[1]{$^{[#1]}$}
\newcommand\avr[1]{\left\langle #1 \right\rangle}
\newcommand\lambdamsb{\Lambda_5^{\rm \sss \overline{MS}}}
\newcommand\qqb{{q\overline{q}}}
\newcommand\qb{\overline{q}}
\newcommand\cs{\hat{s}}
\newcommand\ct{\hat{t}}
\newcommand\cu{\hat{u}}
\newcommand\ycc{y_{c\bar{c}}}
\newcommand\mcc{M_{c\bar{c}}}
\newcommand\mqcc{M^2_{c\bar{c}}}
\newcommand\MSB{{\rm \overline{MS}}}
\newcommand\DIG{{\rm DIS}_\gamma}
\renewcommand\topfraction{1}       
\renewcommand\bottomfraction{1}    
\renewcommand\textfraction{0}      
\setcounter{topnumber}{5}          
\setcounter{bottomnumber}{5}       
\setcounter{totalnumber}{5}        
\setcounter{dbltopnumber}{2}       
\newsavebox\tmpfig
\newcommand\settmpfig[1]{\sbox{\tmpfig}{\mbox{\ref{#1}}}}
\begin{titlepage}
\nopagebreak
\vspace*{-1in}
{\leftskip 11cm
\normalsize
\noindent   
\newline
GEF-TH-4/1996 \newline
ETH-TH/96-08   \newline
{\tt hep-ph/yymmxxx}

}
\vskip 1.0cm
\vfill
\begin{center}
{\large \bf Constraints on the polarized gluon density in the proton from
charm photoproduction}
\vfill
\vskip .6cm 
{\bf Stefano Frixione$^a$}
\vskip .2cm
{ETH, Z\"urich, Switzerland}
\vskip .5cm                                               
{\bf Giovanni Ridolfi}
\vskip .2cm
{INFN, Sezione di Genova, Genoa, Italy.}

\end{center}
\vfill
\nopagebreak
\begin{abstract}
{\small
We consider the possibility of a direct determination of the polarized gluon 
density in the proton using charm production with polarized beams at HERA.
We study total cross sections and distributions at leading order using
different parametrizations of the polarized gluon density.
We conclude that charm photoproduction data can be used
to constrain the polarized gluon density in the proton if an integrated
luminosity of 100~pb$^{-1}$ will be achieved at HERA.
}
\end{abstract}
\vskip0.7truecm
\noindent
~$^a$ Work supported by the National Swiss Foundation
\end{titlepage}

A direct measure of the spin-dependent gluon density in the proton, $\Delta
g(x,Q^2)$, has never been performed (see ref.~[\ref{review}] 
for a recent review
on polarized nucleon structure). Polarized deep inelastic scattering data allow
to extract the structure function $g_1(x,Q^2)$ at different values of $x$ and
$Q^2$ and to obtain an indirect determination of $\Delta g$ through
scaling violation~[\ref{BFR1}-\ref{StratmannVogelsang}]. However, such
indirect determinations are affected by large uncertainties, because of limited
statistics and limited coverage of the $x,Q^2$ range of the presently available
experimental data. 
On the other hand, a precise knowledge of the polarized gluon
density is important in order to understand various aspects of the structure of
polarized nucleons. For example, it has long been known [\ref{AltarelliRoss},
\ref{AltarelliLampe}] that the anomalous gluon contribution to the first moment
of $g_1$, although formally of order $\as$, does not vanish in the large-$Q^2$
limit,
and therefore affects the extraction of the singlet axial charge from polarized
deep inelastic scattering data. Furthermore, an independent determination of
$\Delta g$ would allow to test the reliability of the perturbative expansion,
and to assess the importance of possible non-perturbative contributions. 

In order to measure $\Delta g$ directly, it is necessary to consider processes
which are only or predominantly initiated by gluons; many have been
studied in the past, like for example
heavy quarkonia production in photon-gluon fusion~[\ref{guillet}], or 
the production of large-$k_T$ jet pairs in deep-inelastic
scattering~[\ref{AltarelliStirling},\ref{CarlitzCollinsMueller}]. 
Another interesting possibility is the production of heavy quarks in
photon-proton collisions. This process has already been considered in
refs.~[\ref{GluckReya}-\ref{Vogelsang}] in view of the possibilities of the
electron-proton collider HERA, 
and the conclusion was reached that the study of 
total cross section asymmetries for heavy-quark
photoproduction at HERA can be of little help in constraining the polarized
gluon density in the proton. In this letter we reconsider this problem, in the
light of recent experimental information on polarized deep inelastic scattering
[\ref{SMC},\ref{E143}] and improved theoretical understanding, and of the
planned values of luminosity for the HERA collider.
We will also consider the impact of realistic experimental cuts
on the observed quantities, and discuss some theoretical uncertainties. We will
always refer to the production of charm quark-antiquark pairs, since this is
the most favourable case for the kind of study we are considering here.

The electron-proton cross section for the production of heavy quarks 
can be reliably computed in the Weizs\"acker-Williams
approximation~[\ref{WWpaper}],
where only the contribution of on-shell collinear photons radiated by the
incoming electron beam is retained.
Alternatively, the HERA experiments are capable to tag the scattered electron,
thus determining the energy of the photon that initiated the reaction under 
study. In this kind of analysis, HERA works as a photon-proton collider with
a well defined photon energy, ranging between 70 and 270 GeV approximately.
In both cases, the relevant parton subprocess at the leading order is
\beq
\label{subproc}
g(\hat{p},\lambda_g)+\gamma(\hat{q},\lambda_\gamma)\to c(k)+\bar{c}(k'),
\eeq
where four momenta and helicities are indicated in brackets. The 
${\cal O}(\aem\as)$ partonic cross section for the process (\ref{subproc})
is given by~[\ref{GluckReya},\ref{GluckReyaVogelsang}]
\beq
d\hat{\sigma}_{\gamma g}(\cs,\ct,\lambda_g,\lambda_\gamma)=
\frac{e_c^2\as}{16\cs} 
\left[\Sigma+\lambda_g\lambda_\gamma\Delta
\right] \beta d\cos\theta,
\eeq
where
\beqn
&&\Sigma=-\frac{8 m^4 \cs^2}{\ct^2 \cu^2}+2\frac{\ct^2+\cu^2+4m^2\cs}{\ct\cu},
\\
&&\Delta=
\frac{4m^2(\ct^3+\cu^3)}{\ct^2 \cu^2}+2\frac{\ct^2+\cu^2-2 m^2 \cs}{\ct\cu},
\eeqn
$m$ is the charm quark mass, $e_c$ its electric charge, 
$\beta=\sqrt{1-4m^2/\cs}$ and $\theta$ is the scattering angle in 
the $\gamma g$ center-of-mass frame. We have defined
\beq
\cs=(\hat{p}+\hat{q})^2,\;\;\; \ct=(\hat{p}-k)^2-m^2, \;\;\;
\cu=(\hat{p}-k')^2-m^2.
\eeq
The quantity of interest is the difference
\beq
\label{deltasigma}
d\Delta\sigma_{\gamma p}=
\frac{1}{2}
\left(d\sigma_{\gamma p}^{\uparrow\uparrow}
-d\sigma_{\gamma p}^{\uparrow\downarrow}\right),
\eeq
where $d\sigma_{\gamma p}^{\uparrow\uparrow}$ and 
$d\sigma_{\gamma p}^{\uparrow\downarrow}$ are the differential cross 
sections for the $c\bar{c}$ photoproduction process, 
with parallel and antiparallel polarizations of the incoming photon
and proton, respectively. In fact, it can be easily shown that at leading order
the quantity in eq.~(\ref{deltasigma}) can be written as
\beq
d\Delta\sigma_{\gamma p}(s,t)=
\Delta g(x_g,\muf^2)d\Delta\hat{\sigma}_{\gamma g}(\cs,\ct,\mur^2)dx_g,
\label{sigmagp}
\eeq
where
\beqn
d\Delta\hat{\sigma}_{\gamma g}(\cs,\ct)&=&\frac{1}{4}
\left(
d\hat{\sigma}_{\gamma g}(\cs,\ct,+1,+1)
+d\hat{\sigma}_{\gamma g}(\cs,\ct,-1,-1)\right.
\nonumber\\
&&\phantom{\frac{1}{4}}\left.
-d\hat{\sigma}_{\gamma g}(\cs,\ct,+1,-1)
-d\hat{\sigma}_{\gamma g}(\cs,\ct,-1,+1)
\right).
\eeqn
In eq.~(\ref{sigmagp}) we defined $s=(p+\hat{q})^2$ and $t=(p-k)^2-m^2$, 
where $p$ is the four-momentum of the incoming proton, and $x_g$ is the 
fraction of the proton longitudinal momentum carried by the gluon;
$\mur$ and $\muf$ are the renormalization and factorization scales.

In the case of electroproduction, using the Weizs\"acker-Williams 
approximation we have
\beq
d\Delta\sigma_{ep}(s,t)=\Delta f_{\sss WW}(x_\gamma,Q^2_{\sss WW})
\Delta g(x_g,\muf^2)d\Delta\hat{\sigma}_{\gamma g}(\cs,\ct,\mur^2)
dx_g dx_\gamma\,,
\label{deltaep}
\eeq
where
\beq
\Delta f_{\sss WW}(x_\gamma,Q^2_{\sss WW})=
\frac{\aem}{2\pi}\frac{1-(1-x_\gamma)^2}{x_\gamma}
\log\frac{Q^2_{\sss WW}(1-x_\gamma)}{m^2_ex_\gamma^2},
\eeq
and we defined $s=(p+q)^2$, $q$ being the electron four-momentum.
The mass scale $Q^2_{\sss WW}$ entering the Weizs\"acker-Williams
function has been chosen as discussed in ref.~[\ref{FMNRWW}].

We will consider the three fits to $\Delta g$ presented in 
ref.~[\ref{GehrmannStirling}], which we will indicate with GS-A, GS-B, GS-C,
and those presented in ref.~[\ref{BFR2}], denoted by BFR-AB, BFR-OS and BFR-AR.
The three parametrizations of $\Delta g$ given in ref.~[\ref{BFR2}]
are obtained by performing fits to data within three different subtraction
schemes for collinear divergences. Since a next-to-leading order calculation
of the polarized partonic cross sections is not available, the scheme choice
is immaterial in our analysis.

We begin by considering total cross sections. In the first row of
table~\ref{totals} we present (columns indicated with I) the total charm
production cross section computed at next-to-leading order
[\ref{EllisNason}-\ref{FMNR}]. We show both the electroproduction results,
obtained using the Weizs\"acker-Williams approximation at the HERA
center-of-mass energy of 314 GeV, and the photoproduction results, with a
center-of-mass energy of the photon-proton system of 200 GeV. 
The factorization and renormalization scales are chosen equal to $2m$ and
$m$ respectively (see ref.~[\ref{FMNRphoto}] for a detailed discussion of
scale choices).
\begin{table}
\footnotesize
\begin{center}
\begin{tabular}{|l||c|c|c||c|c|c|} \hline
& \multicolumn{3}{c||}{$e p$}
& \multicolumn{3}{c|}{$\gamma p$} 
\\ \hline
      & I & II & III & I & II & III
\\ \hline \hline 
$\sigma$ $\scriptstyle (NLO)$ $(\mu b)$
& 0.5505 & 0.09653 & 0.04542 & 3.929 & 0.9205 & 0.5268 
\\ \hline
$\Delta\sigma/\sigma$ $\scriptstyle GS-A$
& $7.19\cdot 10^{-4}$ & $5.66\cdot 10^{-3}$ & $9.58\cdot 10^{-3}$ 
& $2.78\cdot 10^{-3}$ & $1.87\cdot 10^{-2}$ & $2.82\cdot 10^{-2}$
\\ \hline
$\Delta\sigma/\sigma$ $\scriptstyle GS-B$
& $5.97\cdot 10^{-4}$ & $5.96\cdot 10^{-3}$ & $9.80\cdot 10^{-3}$ 
& $2.51\cdot 10^{-3}$ & $2.04\cdot 10^{-2}$ & $2.98\cdot 10^{-2}$
\\ \hline
$\Delta\sigma/\sigma$ $\scriptstyle GS-C$
& $3.87\cdot 10^{-5}$ & $3.56\cdot 10^{-3}$ & $5.41\cdot 10^{-3}$ 
& $3.38\cdot 10^{-3}$ & $1.34\cdot 10^{-2}$ & $1.79\cdot 10^{-2}$
\\ \hline
$\Delta\sigma/\sigma$ $\scriptstyle BFR-AB$
& $5.01\cdot 10^{-4}$ & $8.04\cdot 10^{-3}$ & $1.28\cdot 10^{-2}$ 
& $2.03\cdot 10^{-3}$ & $2.87\cdot 10^{-2}$ & $4.02\cdot 10^{-2}$
\\ \hline
$\Delta\sigma/\sigma$ $\scriptstyle BFR-OS$
& $5.71\cdot 10^{-4}$ & $5.56\cdot 10^{-3}$ & $9.20\cdot 10^{-3}$ 
& $2.24\cdot 10^{-3}$ & $1.91\cdot 10^{-2}$ & $2.79\cdot 10^{-2}$
\\ \hline
$\Delta\sigma/\sigma$ $\scriptstyle BFR-AR$
& $5.49\cdot 10^{-4}$ & $6.46\cdot 10^{-3}$ & $1.05\cdot 10^{-2}$ 
& $2.26\cdot 10^{-3}$ & $2.26\cdot 10^{-2}$ & $3.24\cdot 10^{-2}$
\\ \hline
$1/\sqrt{2\sigma{\cal L}}$
& $9.53\cdot 10^{-5}$ & $2.28\cdot 10^{-4}$ & $3.32\cdot 10^{-4}$ 
& $2.77\cdot 10^{-4}$ & $5.73\cdot 10^{-4}$ & $7.57\cdot 10^{-4}$ 
\\ \hline
\end{tabular}
\ccaption{}{\label{totals}
Total cross sections and total cross section asymmetries for $c\bar{c}$
production in $ep$ collisions at $\sqrt{s_{ep}}=314$~GeV and in
$\gamma p$ collisions at $\sqrt{s_{\gamma p}}=200$~GeV, for different
choices of the polarized gluon density. The charm quark mass is 1.5 GeV,
and the unpolarized gluon distribution is MRSA. The integrated luminosity
for the $ep$ system is 100~pb$^{-1}$.
}
\end{center}
\end{table}                                

We also show (columns II) the values of total cross
sections with the conditions
\beqn
\label{cutpt}
&&p_{\sss T} > 2\; {\mathrm GeV}
\\
\label{cuteta}
&&\abs{\eta} < 1.5,
\eeqn
imposed on the transverse momentum $p_{\sss T}$ 
and the pseudorapidity $\eta$ of
the observed heavy quark, (assuming that only one of the two heavy quarks
produced is fully reconstructed). The conditions (\ref{cutpt}) and
(\ref{cuteta}) approximately reproduce 
the present experimental situation of the HERA experiments.
Finally, we show (columns III) the effect of applying the Peterson
fragmentation function~[\ref{Peterson}] to the produced charm quarks, 
in order to estimate the impact of hadronization phenomena. In columns III 
the cuts of eqs.~(\ref{cutpt}) and (\ref{cuteta}) are still applied.
Notice that in the absence of the cuts
the total cross section values would be insensitive to fragmentation
effects. On the other hand, the cuts introduce in the total cross section 
a dependence upon the $p_{\sss T}$ and $\eta$ distributions;
in particular, since Peterson fragmentation softens the $p_{\sss T}$
spectrum, the low-$p_{\sss T}$ region gives a contribution to the
cross section larger than in the bare quark case. This explains the lower
values of columns III with respect to columns II.

In the following six rows we display the values of the asymmetry
$\Delta\sigma/\sigma$ in the same cases, obtained with the six different
parametrizations of $\Delta g$. It must be stressed that the unpolarized cross
section $\sigma$ that appears in the denominator of the asymmetry is computed
at the leading order, because a next-to-leading order calculation for the
polarized cross section is not available. However, one might expect that the
effect of radiative corrections approximately cancels in the ratio
$\Delta\sigma/\sigma$. Notice that the asymmetries obtained with the
fragmented quarks are larger than those for bare quarks, at variance
with the case of the total cross sections. This is due to the fact that
the dominant contribution to $\Delta\sigma/\sigma$ is given by the 
large-$p_{\sss T}$ region, where the asymmetry values are larger for
fragmented quarks than for bare ones. We will discuss again this issue
when dealing with differential cross sections.

The next-to-leading order value of $\sigma$ can then be
used to estimate the sensitivity of the experiment. A rough estimate of the 
minimum value of the asymmetry observable at HERA can be obtained by requiring 
the difference between the numbers of events with parallel and antiparallel
polarizations of the initial state particles to be larger than the 
statistical error on the total number of observed events. This gives
\beq
\left[\frac{\Delta\sigma}{\sigma}\right]_{min}\simeq
\frac{1}{\sqrt{2\sigma{\cal L}\epsilon}},
\label{minassmtr}
\eeq
where $\cal L$ is the integrated luminosity and the factor $\epsilon$
accounts for the experimental efficiency for charm identification
and the fact that the initial beams are not completely polarized.
The values of $1/\sqrt{2\sigma{\cal L}}$ in the various cases are given in the 
last row of table~\ref{totals}, for ${\cal L}=100$~pb$^{-1}$. 
In the case of photoproduction, the effective luminosity has been
estimated by assuming that the results obtained with photons such
that 170~GeV$<\sqrt{s_{\gamma p}}<$~230 GeV can be reliably described
by the cross section for an incoming monochromatic photon
with $\sqrt{s_{\gamma p}}=200~GeV$. The $ep$ luminosity of 100~pb$^{-1}$
has then been rescaled by a factor equal to the integral of the
Weizs\"acker-Williams function in the appropriate range.

By inspection of table~\ref{totals}, we conclude that the asymmetries
of practical interest (columns II and III)
are always larger than the corresponding minimum observable values with
an integrated luminosity of 100~pb$^{-1}$ even if an experimental efficiency
of 1\% is assumed. We also observe that the asymmetry is generally larger
in the phase space region defined by eqs.~(\ref{cutpt}) and (\ref{cuteta}).
In practice, the quantity ${\cal L}\Delta\sigma$ is required to be larger
than 5 or 10 events, in order to distinguish the signal from the background.
This condition appears to be fulfilled for all the cases considered
in table~\ref{totals}, with ${\cal L}=100^{-1}$.
We also observe that all the chosen parametrizations for the polarized gluon
density give similar results for the total cross section asymmetry. It seems 
therefore difficult to distinguish among them with this kind of measurement.

We now turn to differential distributions. From table~\ref{totals} we know that
the asymmetry tends to become larger in the phase-space region where the
conditions (\ref{cutpt}) and (\ref{cuteta}) are satisfied; we will therefore 
present our predictions with the kinematical cuts (\ref{cutpt}) and 
(\ref{cuteta}) applied. We will consider only the GS-A and BFR-AB
parametrizations. 

\begin{figure}[ptbh]
  \begin{center}
    \mbox{
      \epsfig{file=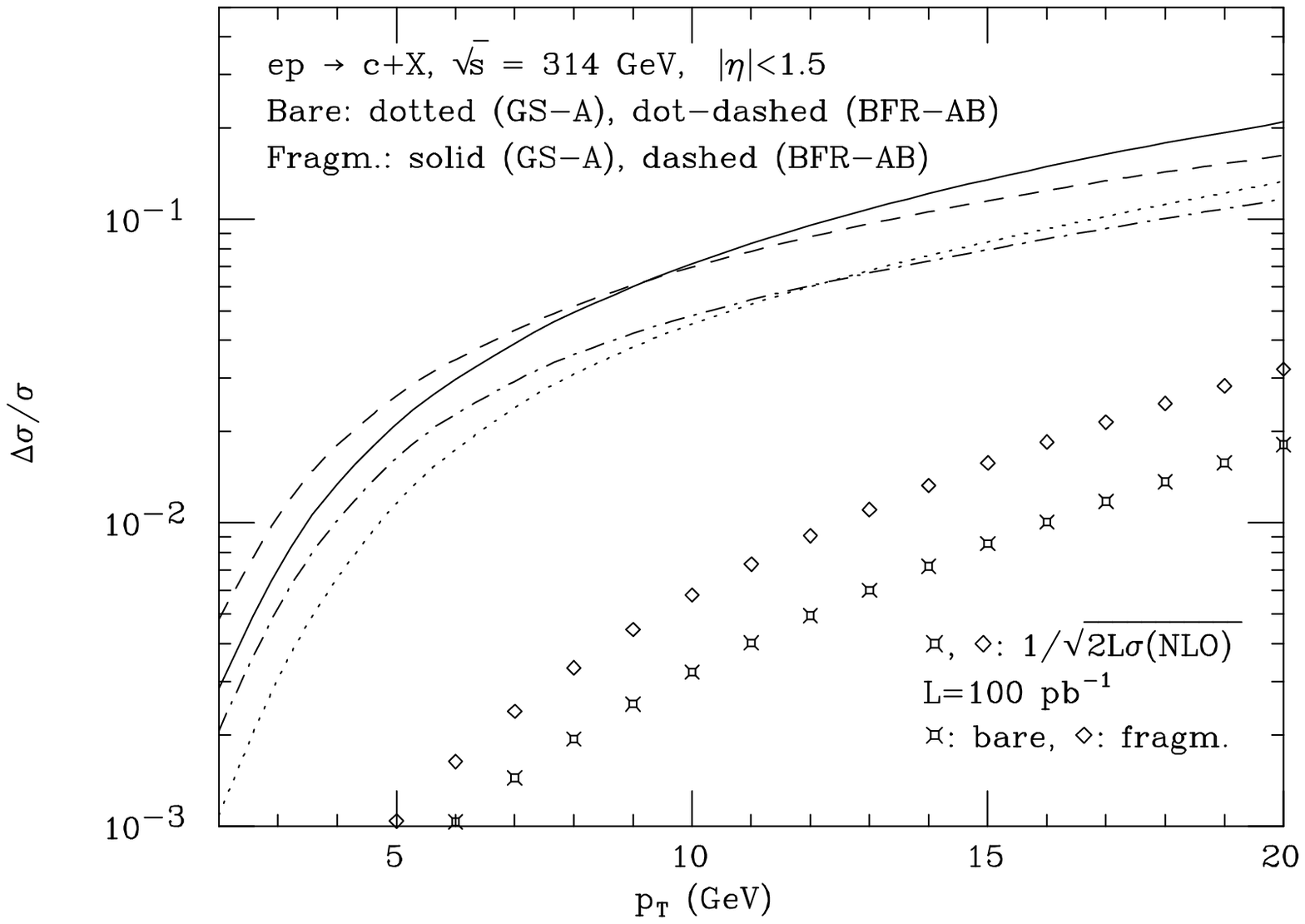,width=0.80\textwidth}
      }
  \ccaption{}{\label{ptel}
Asymmetry cross section versus transverse momentum in $ep$ collisions at
$\sqrt{s}=314$~GeV. The minimum observable asymmetry, computed
at next-to-leading order, is also displayed.
}
  \end{center}
\end{figure}

In fig.~\ref{ptel} we show the asymmetry versus the transverse momentum of
the observed quark for $ep$ collision at a center-of-mass 
energy of 314~GeV. The scale choice is $\mur=\muo$, $\muf=2\muo$, with
$\muo=\sqrt{p_{\sss T}^2+m^2}$.
As previously observed, the asymmetry increases with increasing transverse
momentum. Observe also that the asymmetry for fragmented quarks 
is always larger than
that for bare quarks over the whole $p_{\sss T}$ range considered, consistently
with the analogous behaviour of total cross sections. The results obtained with
the two parametrizations for $\Delta g$ are quite close to each other, although
they display slightly different shapes: the BFR parametrization produces larger
asymmetries in the low-$p_{\sss T}$ region. 
The effect of Peterson fragmentation on the shape of this 
distribution is very small. In fig.~\ref{ptel} we also show the quantity
defined in eq.~(\ref{minassmtr}) for $\epsilon=1$ and ${\cal L}=100$~pb$^{-1}$,
as an indication for the minimum observable value of the asymmetry. In this
case, the quantity $\sigma$ appearing 
in the RHS of eq.~(\ref{minassmtr}) is the
next-to-leading order cross section for a given $p_{\sss T}$ bin. We used a bin
size of 1 GeV; clearly, a larger bin size would decrease the minimum observable
asymmetry; on the other hand, by enlarging the bins the resolution of the
measurement would get worse. From fig.~\ref{ptel}, we can see that there exists
a wide $p_{\sss T}$ region where the predicted asymmetries are more than one
order of magnitude above the corresponding minimum observable asymmetry.
Furthermore, the value of the asymmetry for $p_{\sss T}$ above 10~GeV
is of the order of few tens of percent.
\begin{figure}[ptbh]
  \begin{center}
    \mbox{
      \epsfig{file=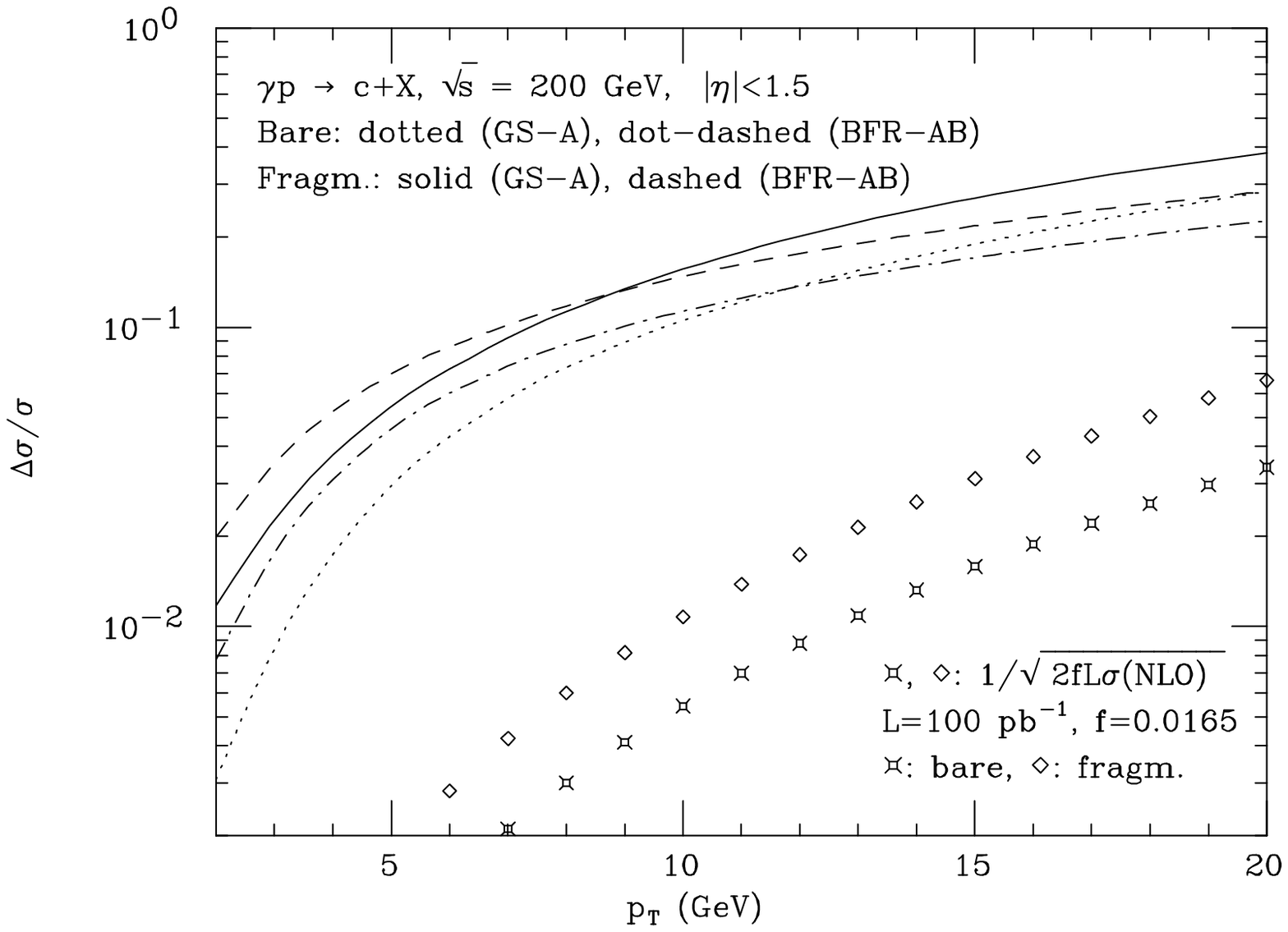,width=0.80\textwidth}
      }
  \ccaption{}{\label{ptph}
Asymmetry cross section versus transverse momentum in $\gamma p$ collisions at
$\sqrt{s}=200$~GeV. The minimum observable asymmetry, computed
at next-to-leading order, is also displayed.
}
  \end{center}
\end{figure}
We can
therefore conclude that the $p_{\sss T}$ distribution could be used to extract
information on the polarized gluon density, even with a low experimental
efficiency. The shape of the distribution could in principle provide with the
possibility of discriminating  among the various sets for $\Delta g$. However,
since the predictions for different sets are quite close to each other, this
kind of measurement does not appear to be feasible with the planned luminosity
values. 

Figure~\ref{ptph} is the analogous of fig.~\ref{ptel} for $\gamma p$ collisions
at a center-of-mass energy of 200 GeV. The asymmetry values turn out to be
larger than for $ep$ collisions in the whole $p_{\sss T}$ range, and the curves
appear to be slightly flatter than in the previous case. On the other hand, the
minimum observable value is larger than before. We computed the effective
luminosity for the incoming photon as in the case of total cross sections. 

It is also interesting to consider fully exclusive distributions, despite the 
fact that a large statistics is needed in order to reconstruct completely 
both the produced charm and anti-charm. 
In analogy with the unpolarized case (see ref.~[\ref{FMNRglu}]), we can rewrite
eq.~(\ref{deltaep}) as
\beq
\frac{d\Delta\sigma_{ep}}{d\ycc d\mqcc}=\frac{1}{s}
\Delta f_{\sss WW}(x_\gamma,Q^2_{\sss WW})
\Delta g(x_g,\muf^2)\Delta\hat{\sigma}_{\gamma g}(\mqcc),
\label{ymdistr}
\eeq
with
\beqn
x_\gamma&=&\frac{\mcc}{\sqrt{s}}\exp(-\ycc),
\label{xgamma}
\\
x_g&=&\frac{\mcc}{\sqrt{s}}\exp(\ycc).
\label{xg}
\eeqn
In eqs.~(\ref{ymdistr}), (\ref{xgamma}) and~(\ref{xg}) 
$\mcc^2=\cs$ is the invariant mass of the $c\bar{c}$ pair, 
and $\ycc$ its rapidity in the electron-proton center-of-mass frame. 
\begin{figure}[ptbh]
  \begin{center}
    \mbox{
      \epsfig{file=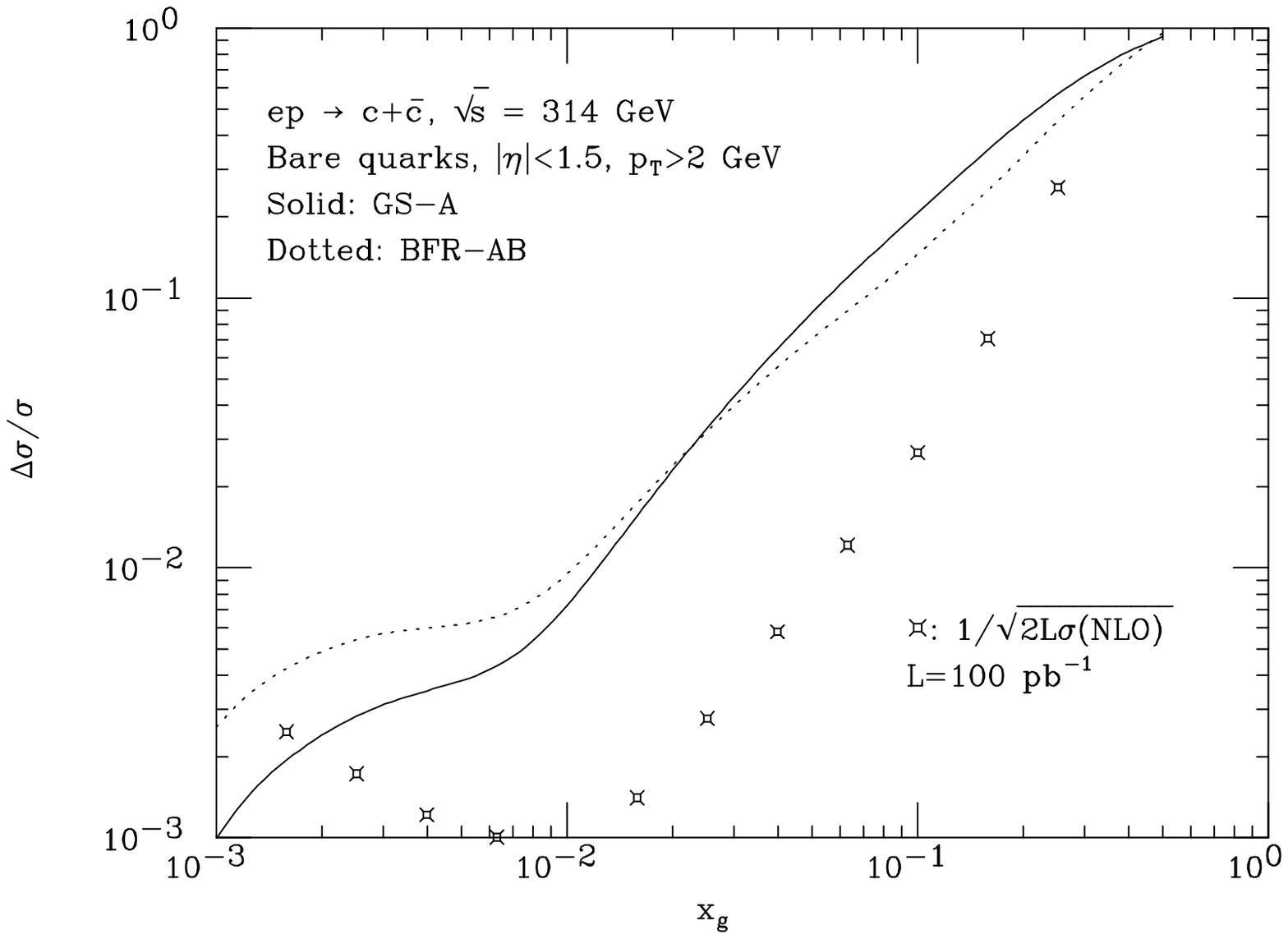,width=0.80\textwidth}
      }
  \ccaption{}{\label{xgel}
Asymmetry cross section versus $x_g$ in $ep$ collisions at
$\sqrt{s_{ep}}=314$~GeV. The minimum observable asymmetry, computed
at next-to-leading order, is also displayed.
}
  \end{center}
\end{figure}
We have positive rapidities in the proton direction.
Identifying the LHS of eq.~(\ref{ymdistr})
with the experimental data, we can invert this equation to get
\beq
\Delta g(x_g,\muf^2)=
\frac{s}{\Delta f_{\sss WW}(x_\gamma,Q^2_{\sss WW})
\Delta\hat{\sigma}_{\gamma g}(\mqcc)}
\left(\frac{d\Delta\sigma_{ep}}{d\ycc d\mqcc}\right)^{data}.
\label{Dgvsdata}
\eeq

We present our results directly in terms of the variable $x_g$,
exploiting the identity
\beq
\frac{d\Delta\sigma_{ep}}{d\ycc d\mqcc}=
x_g\,\frac{d\Delta\sigma_{ep}}{dx_g d\mqcc}\,.
\eeq
In fig.~\ref{xgel} we show our predictions for the $x_g$ asymmetry in $ep$
collisions at a center-of-mass energy of 314 GeV. In this case, the reference
scale $\muo$ is equal to $\sqrt{(p_{c\sss T}^2+p_{\bar{c}\sss T}^2)/2+m^2}$. 
We do not present the result for the fragmented
quarks. In fact, when Peterson fragmentation is applied, $x_g$
is not expressed in terms of the invariant mass and rapidity of the pair
as in eq.~(\ref{xg}). 
The correct relationship could in principle be worked out,
but it is more straightforward to obtain the $c\bar{c}$ cross 
section by directly performing the deconvolution on the measured cross section
for charmed hadron production. This procedure can be applied
using different hadronization models, thus obtaining an
estimate of the dependence of the result upon the assumed
hadronization mechanism.

By inspection of fig.~\ref{xgel}, we can conclude that
the polarized gluon density can be probed with a good resolution (we used a
bin size of 0.2 in the variable $\log x_g$) in the range $(10^{-2},10^{-1})$,
if the experimental efficiency is not too small. The GS-A and
BFR-AB results are quite similar, and it appears unlikely that this kind of
measurement could discriminate between them. 

In this paper we have studied the problem of measuring the polarized
gluon density in the proton using the charm data which will be 
collected at the HERA collider in the polarized configuration. We used a
leading-order QCD calculation, since the polarized short distance
partonic cross sections are not available at the next-to-leading order.
For this reason, we presented our results in terms of asymmetries,
which are more likely to be stable under radiative corrections than
the polarized or unpolarized cross sections. We have shown that,
assuming an integrated luminosity of 100 pb$^{-1}$, the total cross
section asymmetries can be measured even with a very low experimental
efficiency. The measurement of single-inclusive and double-differential
distributions also appears to be feasible, although in the latter case
a higher luminosity is certainly needed.
With a luminosity of 100 pb$^{-1}$, it seems difficult to distinguish
among the various parametrizations for the polarized gluon density.

We have not considered the inclusion of the hadronic component of the
photon, which is known to be important in the HERA regime. Unfortunately, no
information on the polarized parton densities in the photon is available up to
now, and therefore no theoretical estimate can be given. It is
known~[\ref{FNRdiff}] that the $p_{\sss T}$ spectrum of the 
hadronic component is softer than the pointlike one
in the unpolarized case, and therefore the cut of
eq.~(\ref{cutpt}) reduces the impact of the hadronic component on the full
unpolarized results. This might not happen in the polarized case. However, 
for hadronic photons a sizeable fraction of
the photon momentum is lost into hadronic fragments, that can be observed
in the photon direction. In this way it should be possible to disentangle the 
hadronic component from the pointlike one.

\section*{Acknowledgements}
We wish to thank S.~Forte, C.~Grab, P.~Nason and L.~Rolandi for useful
suggestions.

\newpage
\begin{reflist}
\item\label{review}
   S.~Forte, Proceedings of the 6th International Conference on 
   Elastic and Diffractive Scattering, Blois, France, 20-24 Jun 1995,
   preprint CERN-TH-95-305, hep-ph/9511345;\\
   G.~Altarelli, Proceedings of the International School of Subnuclear
   Physics, Erice, 1989, preprint CERN-TH-5675-90.
\item\label{BFR1}
   R.D.~Ball, S.~Forte and G.~Ridolfi, \np{B444}{95}{287}.
\item\label{GehrmannStirling}
   T. Gehrmann and W.J.~Stirling, preprint DTP/95/82, hep-ph/9512406,
   to appear in Phys. Rev. D.
\item\label{BFR2}
   R.D.~Ball, S.~Forte and G.~Ridolfi, hep-ph/9510449, 
   to appear in Phys.~Lett.~B.
\item\label{StratmannVogelsang}
   M.~Gluck, E.~Reya, M.~Stratmann and W.~Vogelsang, preprint DO-TH-95-13,
   hep-ph/950834, to appear in Phys. Rev. D.
\item\label{AltarelliRoss}
   G.~Altarelli and G.G.~Ross, \pl{B212}{88}{391}.
\item\label{AltarelliLampe}
   G.~Altarelli and B.~Lampe, \zp{C47}{90}315.
\item\label{guillet}
   J.-P.~ Guillet, \zp{C39}{88}{75}.
\item\label{AltarelliStirling}
   G.~Altarelli and W.J.~Stirling, {\it Particle World} {\bf 1}(1989)40.
\item\label{CarlitzCollinsMueller}
   R.D.~Carlitz, J.C.~Collins and A.H.~Mueller, \pl{B214}{88}{229}.
\item\label{GluckReya}
   M.~Gl\"uck and E.~Reya, \zp{C39}{88}{569}.
\item\label{GluckReyaVogelsang}
   M.~Gl\"uck, E.~Reya and W.~Vogelsang, \np{B351}{91}{579}.
\item\label{Vogelsang}
   W.~Vogelsang, in Physics at HERA, Proc. of the Workshop Vol. 1,
   eds. W.~B\"uchmuller and G.~Ingelman (1991).
\item\label{SMC}
   D.~Adams et al., SMC Collaboration, \pl{B329}{94}{399};\\
   \pl{B357}{95}{248}.
\item\label{E143}
   K.~Abe et al., E143 Collaboration, \prl{74}{95}{346};\\
   \prl{75}{95}{25}.
\item\label{WWpaper}
   C.F.~Weizs\"acker, \zp{88}{34}{612};\\
   E.J.~Williams, \pr{45}{34}{729}.
\item\label{FMNRWW}
   S.~Frixione, M.L.~Mangano, P.~Nason and G.~Ridolfi, \pl{B319}{93}{339}.
\item\label{EllisNason}
   R.K.~Ellis and P.~Nason, \np{B312}{89}{551}.
\item\label{SmithNeerven}
   J.~Smith and W.L.~van Neerven, \np{B374}{92}{36}.
\item\label{FMNR}
   S.~Frixione, M.L.~Mangano, P.~Nason and G.~Ridolfi, \pl{B348}{95}{633}.
\item\label{FMNRphoto}
   S.~Frixione, M.L.~Mangano, P.~Nason and G.~Ridolfi, \np{B412}{94}{225}.
\item\label{Peterson}
   C.~Peterson, D.~Schlatter, I.~Schmitt and P.~Zerwas, \pr{D27}{83}{105}.
\item\label{FMNRglu}
   S.~Frixione, M.L.~Mangano, P.~Nason and G.~Ridolfi, \pl{B308}{93}{137}.
\item\label{FNRdiff}
   S.~Frixione, P.~Nason and G.~Ridolfi, \np{B454}{95}{3}.
\end{reflist}
\end{document}